# Inverse design of artificial skins

Zhiguang Liu[1,2,8], Minkun Cai[1,8], Shenda Hong[3,8], Junli Shi[1,8], Sai Xie[1], Chang Liu[2], Huifeng Du[2], James D. Morin[2], Gang Li[1], Wang Liu[7], Hong Wang[1], Ke Tang[6], Nicholas X. Fang[2,5]*, Chuan Fei Guo[1,4]*

**Affiliations:**

[1] Department of Materials Science and Engineering, Southern University of Science and Technology, Shenzhen 518055, China

[2] Department of Mechanical Engineering, Massachusetts Institute of Technology, Cambridge, MA 02139, USA

[3] National Institute of Health Data Science, Peking University, Beijing 100191, China

[4] Centers for Mechanical Engineering Research and Education at MIT and SUSTech, Southern University of Science and Technology, Shenzhen 518055, China

[5] Department of Mechanical Engineering, The University of Hong Kong, Hong Kong 999077, China

[6] Guangdong Provincial Key Laboratory of Brain-inspired Intelligent Computation, Department of Computer Science and Engineering, Southern University of Science and Technology, Shenzhen 518055, China

[7] Department of Modern Mechanics, University of Science and Technology of China, Hefei 231600, China.

[8] These authors contributed equally: Zhiguang Liu, Minkun Cai, Shenda Hong, Junli Shi.

*Correspondence should be addressed to N.X.F. (Email: nicfang@mit.edu) or C.F.G. (Email: guocf@sustech.edu.cn)

**Mimicking the perceptual functions of human cutaneous mechanoreceptors, artificial skins or flexible pressure sensors can transduce tactile stimuli to quantitative electrical signals[1-3]. Conventional methods to design such devices follow a forward structure-to-property routine based**

on trial-and-error experiments/simulations, which take months or longer to determine one solution valid for one specific material. Target-oriented inverse design that shows far higher output efficiency has proven effective in other fields[4,5], but is still absent for artificial skins because of the difficulties in acquiring big data. Here, we report a property-to-structure inverse design of artificial skins based on small dataset machine learning, exhibiting a comprehensive efficiency at least four orders of magnitude higher than the conventional routine. The inverse routine can predict hundreds of solutions that overcome the intrinsic signal saturation problem for linear response in hours, and the solutions are valid to a variety of materials. Our results demonstrate that the inverse design allowed by small dataset is an efficient and powerful tool to target multifarious applications of artificial skins, which can potentially advance the fields of intelligent robots, advanced healthcare, and human-machine interfaces.

**Main Text:**

Tactile sensation is one of the most important senses for humans to interact with their surroundings enabled by cutaneous mechanoreceptors[6]. Artificial skins or flexible pressure sensor arrays mimicking mechanoreceptors to collect haptic information are developing rapidly and have extensively promoted the fields of robotics[7,8], health monitoring[9,10], human-machine interfaces[11,12], and Metaverse[13,14]. Conventional methods to design and fabricate flexible pressure sensors are based on a forward routine— empirically exploring a synergistic effect of the materials and structures on the properties of sensors using an uncertain trial-and-error process (Fig. 1a), mostly by carrying out a considerable number of experiments. The conventional routine, however, often takes months or even years to determine one optimized device structure, while the structure is often effective to only one specific material, and deviation in properties may still occur (Fig. 1b)[15,16]. Furthermore, the conventional routine becomes even more inefficiency in dealing with complex problems, such as designing sensors with combined properties (sensitivity, linearity, working range, etc.).

Inverse design is a routine that reverses the conventional design process, targeting at desired outputs to predict input parameters. Machine learning (ML)-based inverse design has been well performed in mechanics, photonics, and materials science[17-19], but still lacks successful cases in flexible sensors. The primary reason is that the multidisciplinary correlation (material-microstructure-property) of the device complicates the problem and thus boosts the needed amount of data (thousands or even millions) to explore the whole design scope, while the experimental and/or simulated data acquisition is time-consuming and has low reliability (Fig. 1b). Therefore, an urgent need to advance the field of flexible electronics is reducing the size of dataset to a minimized number for model construction in target-oriented inverse design.

Here, we report a high-efficiency ML-based inverse routine to design flexible pressure sensors with targeted properties. A model is constructed based on a small dataset of only 100 cases. We use a two-pronged approach to minimize the requirement of data (Fig. 1c): we restrict the design scope by applying a reduced-order model to analytically determine internal constraints without any costly experiments and simulations; and we improve the efficiency of data-collection by proposing a "jumping-selection" method in combination with an iterative surrogate model to provide six times higher efficiency than conventional random-searching. Hundreds of solutions that are effective to a variety of materials, were successfully predicted in hours, in comparison to the conventional routine that takes months or years to determine one solution for one material.

In this study, we verify the effectiveness of the ML-based inverse routine by utilizing the routine to predict a collection of structures capable of addressing the issue of nonlinear response in flexible pressure sensors. The nonlinearity stems from the intrinsic stiffening effect of soft materials, which is associated with their incompressibility (Extended Data Fig. 1). Such stiffening effect leads to signal saturation (i.e., nonlinear response) under increasing pressure and thus an accompanying narrow working range[20,21]. In industry, the nonlinear signal often requires an extra calibration, resulting in increased complexity in both the circuit and algorithm of the sensory system[2,22]. There have been only a few studies working on microstructures and materials that can fight against signal saturation, but all based on conventional forward design[23-27]. In this work, we employ a pillar-like protrusion (a commonly used geometry in flexible pressure sensors[28-31]) as the microstructure model (Fig. 1d). Geometric parameters, radii $x_1$, $y_1$ and $x_2$, $y_2$ of the two ends, and the length $l$, are used to define the profile of this microstructure. We introduce the iontronic sensing mechanism for which capacitive signal magnitude ($C$) is in direct proportion to the interfacial contact area ($A$) between an electrode and a microstructured ionic layer, say, $C\sim A$, to enable a contact area-governed modulation (Extended Data Fig. 2). This simple contact-area model make our method valid to a variety of ionic materials.

In the construction of a ML model, it is expected that the dataset to have cases from every corner of design scope, thus one effect way to minimize the dataset is refining the design scope by internal constraints. As shown in Fig. 2a, the top surface of the microstructure evolves into a flat contact region when loaded, with an area close to its horizontal cross-section area (*CSA*). Therefore, we treat *CSA* as a reduced-order model of *A* to probe the structural deformation under pressure and to determine internal constraints among geometric parameters.

Although *CSAs* of different microstructures are not completely overlapped with their *A* counterparts, they have similar growth trends with respect to the increase of compressive distance (*d*) (Fig. 2b). Such a trend can be represented by the convexity of the *CSA-d* curve, defined as (Supplementary Fig. 1):

$$\text{Convexity} = (CSA_{y_1/2} + CSA_0) - 2 \cdot CSA_{y_1/4} \tag{1}$$

which is originated from the definition of convex function in math[32]. Figure 2c demonstrates the convexity map by sweeping geometric parameters $x_1/x_2$ and $y_1/y_2$. Three representative cases of negative, near-zero, and positive convexity are labeled with A, B, and C, respectively, and their *CSA-d* curves are shown in Fig. 2b. Specifically, the positive convexity curve (case C) reveals a *CSA* growth that closes to the resistance to pressure due to the stiffing effect. Figure 2d shows the averaged value of convexity with respect to the *x* and *y* axes of map. The curves show that the cases with a positive convexity mainly locate at the region $y_1>y_2$ and $x_1<x_2$, which are adopted as the boundaries to reasonably restrict the design scope to 1/4 of the original. In addition, another internal constraint, $y/x≤4$, is applied to avoid buckling that introduces a sudden change in contact area (Extended Data Fig. 3).

An efficient data-collection approach that we call "jumping-selection" further allows to learn the input-output relationship with a reduced size of dataset. A randomly-selected initial dataset (Fig. 2e, left-top) constructs a surrogate model, which is a ML method used to generate estimated data as the surrogate of real-data that is otherwise difficult to acquire[33]. This model rapidly estimates the properties of one-million geometries in seconds, followed by the "jumping-selection" to pick geometries as the supplementary of dataset to enhance the data diversity (see *Methods* for details, Supplementary Text 1 and Fig. 2). Figure 2e shows the datasets picked by the jumping-selection and by the conventional random-selection, respectively, for comparison. Geometries are divided into eligible ($R^2$>0.995) and ineligible ($R^2$<0.995) groups, where $R^2$ is the linear-fitting indicator over a concerned pressure range (0-300 kPa), which fully covers the needs in applications of dexterous manipulation and health monitoring[22,34]. Remarkably, a positive correlation between linearity and convexity is observed (left-bottom part of Fig. 2e and Supplementary Fig. 3), evidencing the effectiveness of the reduced-order model. After five iterations (total 100 cases) that takes a few hours, the percentage of eligible geometries significantly increases from 15% to 45% because the jumping-selection can globally screen out eligible cases. When the threshold is further tightened to 0.999, only two geometries are eligible in the random dataset, which are insufficient to establish an effective model. By contrast, a six times higher efficiency (twelve cases) is found in the jumping-selection dataset (Supplementary Fig. 4). Four out of the twelve cases that present linear response are further verified in experiments (Supplementary Fig. 5).

Based on the selected small dataset, we build a ML model that estimates the relationship between input geometric parameters ($x_1$, $y_1$, $x_2$, $y_2$) and output sensing property, further inversely predicts microstructures that are possibly to present linear response. Figure 2f shows the simulated compressive properties of a representative microstructure (a top selection in the one-million cases) from the jumping-selection dataset and its counterpart from the random dataset. Although in both cases the resistance to pressure increases with displacement due to the stiffing effect, the one from the jumping-selection dataset presents a growth trend coincident with the contact area-displacement curve (red lines), leading to an ultrahigh linearity ($R^2$=0.999). In contrast, the mismatch remains unresolved for the microstructure from the random dataset (blue lines) because the model is misled by the few eligible cases, and thus the structure still exhibits a saturated nonlinear response.

We experimentally explore the contact area evolution of the microstructured interface through electro-etching. The microstructures of the ionic layer are prepared by templating a 3D-printed epoxy mold (Supplementary Fig. 6) with high fidelity of the shape (Supplementary Fig. 7 and Fig. 8). The setup consists of a layer of copper film deposited on silicon wafer serving as the anode, a flat metal indenter as the cathode, and a microstructured layer of ionic gel (polyvinyl alcohol-phosphoric acid, PVA-$H_3PO_4$) placed in between the anode and cathode as the electrolyte (Fig. 2g). By loading pressure and applying a voltage (2 V) on the copper film, only the region that contacts with the microstructured electrolyte is oxidized and changes its color (Supplementary Text 2, Fig. 9 and Fig. 10). Figure 2g illustrates etched patterns at specific pressures (50, 100, 200, and 300 kPa) of the two microstructures that have linear and nonlinear responses (Fig. 2f), which fit well with the simulations (Supplementary Fig. 11).

The gradients of width ($x$) and height ($y$) are key parameters to the contact area modulation. Specifically, the height gradient of the microstructure controls the length of the contact region, that is, only the portion higher than the loading plane will make contact with the top electrode. The width gradient determines the contribution weight of each contact segment to the overall area. Once these two gradients are engineered as the ML model predicted, a pear-shaped 2D expansion of contact area is observed, as marked by the white dashed ellipses in Fig. 2g. The contact initiates from the head side slowly due to a small weight (width). As the compressive distance increases, the stiffening effect appears in such a stable microstructure and endows a positive convexity trend to pressure, while the newly joined contact regions with a larger weight allow the overall contact area to catch up with the pressure, compensating the signal saturation and leading to a linear response. A microstructure without such synergetic gradients will fail to compensate the mismatch. When there is no expansion in the longitudinal direction (white triangles in Fig.

2g), the width gradient does not change the growth trend of contact area. In the case of lacking width gradient, the increasing contact length compensates the mismatch to some extent but lags behind the pressure increment, failing to achieve linear response as well (Extended Data Fig. 4).

The inverse design presents high efficiency in predicting geometries for linear response in iontronic sensors. Figure 3a shows a broad distribution of top 1000 predicted solutions with targeted properties. Six of the 1000 solutions are selected and printed to verify the validity of the structures (each two from the top 100, top 300 and top 1000). All iontronic sensors using the microstructures exhibit ultrahigh linearity ($R^2$=0.999) and high sensitivity (1.78-47.37 pF·kPa$^{-1}$) over a wide pressure range (0-300 kPa).

We further confirmed the linear response under dynamic and repeated loadings. Figure 3b exhibits the capacitance signal under a few sets of repeated loading-unloading cycles with a stepwise increment of ~56 kPa between different sets of loadings. The pressure increment causes a stable capacitance increase of ~0.3 nF, indicating the linearity is maintained during the dynamic loading. The capability to resolve tiny pressure changes at a high basic pressure (0.2 kPa at 200 kPa, or 0.1%) is also confirmed (Supplementary Fig. 12). Furthermore, the linear characteristic of the signal is maintained at different test frequencies (1~100 kHz), as indicated in Fig. 3c.

Our solutions are effective when applied to a diversity of materials that have the contact area-governed modulation. As shown in Fig. 3d, high linearity values of $R^2$>0.996 in sensors are achieved (which is lower than the designed value yet acceptable) when a structure is applied to six different ionic materials, including poly(4-styrenesulfonic acid) (PSS), poly(sodium 4-styrenesulfonate) (PSSS), poly(diallyldimethylammonium chloride) (PDADMAC), polyvinylpyrrolidone (PVP), methyl cellulose (MC), and hydroxypropylmethylcellulose (HPMC) (Supplementary Table 1). We reasonably conclude that the paradigm of the inverse design can provide wider options in both microstructures and materials.

Our flexible sensors can provide contact information between two surfaces, such as real-time monitoring of pressure distribution between the femur and the tibia in TKA. Compared with commercial rigid sensors that can hardly be deployed on curved surfaces[35,36], flexible sensors show better conformability and allow for *in situ* monitoring of pressure distribution, thus can help surgeons evaluate intercompartmental loads and correct assembly imbalances. A pressure sensor array with microstructures designed using the inverse route is applied for TKA pressure mapping (Fig. 4a). We select an ionic gel with a high Young's modulus (~96 MPa, Supplementary Fig. 13) as the ionic layer in our sensors to widen the linear range (Fig. 4b). The sensors are strain-insensitive by using serpentine interconnects—the capacitance to pressure response remains unchanged when the sensors are stretched to 30% strain (Figs. 4b and c), which allows accurate pressure sensing on curved surfaces without the need to decouple strain-induced and pressure-induced signals.

We use two 4×4 sensor arrays to map the pressure distribution between the pseudo femur and the tibia at the medial and lateral compartments (Extended Data Fig. 5). The sensing units in the arrays show little deviation in sensing properties (Supplementary Fig. 14). For each compartment, the wide linear range can help measure loads up to ~50 lbs and fulfills the working range requirement in TKA[35,36]. When the femur is loaded vertically and contacts with both compartments, balanced pressures are observed for different loads from 10 to 40 N (Fig. 4d and Supplementary Fig. 15). The measured load ($F_M$), which is defined as the sum of the products of measured pressure and sensing area for all sensing units, is close to the real applied load ($F_A$) (Fig. 4e), verifying the high accuracy of measurement.

Furthermore, the sensor array can be used to judge different contact statuses. As shown in Fig. 4f, when the femur is deflected to different angles, the pressures remain balanced but the contact point slightly changes. When the femur has an improper oblique angle with the tibia, significant imbalanced pressure distributions are exhibited (Figs. 4g and h). Such real-time and precise pressure mapping can help surgeons correct the contact status during TKA operation and further avoid possible wear or hurt caused by improper assembly.

In summary, we have developed a high-throughput inverse routine based on machine learning for the design of flexible pressure sensors. Our method can construct a small dataset-based model using in hours and predict hundreds of solutions with targeted linear response over a wide pressure range in one second. The efficiency of the inverse design is at least four orders of magnitude higher than that of the conventional trial-and-error forward design. The method also presents high generality in materials, and its validity is verified in all seven materials tested. The use of inverse design may be used for the customization of other properties (not only to fight against signal saturation) in artificial skins, which potentially advances the community of intelligent robots, advanced healthcare, and human-machine interfaces. We expect that the inverse design to be extended to the design of other types of devices.

**Acknowledgments:**

We thank the support by the Centers for Mechanical Engineering Research and Education at MIT and SUSTech (MechERE Centers at MIT and SUSTech).

**Funding:** The work was supported by the National Natural Science Foundation of China (No. T2225017, 52073138, and 62102008), the "Science Technology and Innovation Committee of Shenzhen Municipality" (Grant No. CYJ20210324120202007), the Shenzhen Sci-Tech Fund (No. KYTDPT20181011104007), the "Guangdong Provincial Key Laboratory Program" (No. 2021B1212040001), and the Centers for Mechanical Engineering Research and Education at MIT and SUSTech.

**Author contributions:** Z.L., N.X.F., and C.F.G. conceived the concept. Z.L. designed the microstructures. Z.L. and J.D.M. conducted FEA. Z.L, S.H. and K.T. conducted the ML method. Z.L., M.C., J.S., S.X., and G.L. conducted the experiments. J.S. and S.X. prepared the array and conducted the pressure mapping. Z.L. wrote the paper. C.F.G. and N.X.F. revised the manuscript. All authors reviewed and commented on the manuscript.

**Competing interests:** Authors declare that they have no competing interests.

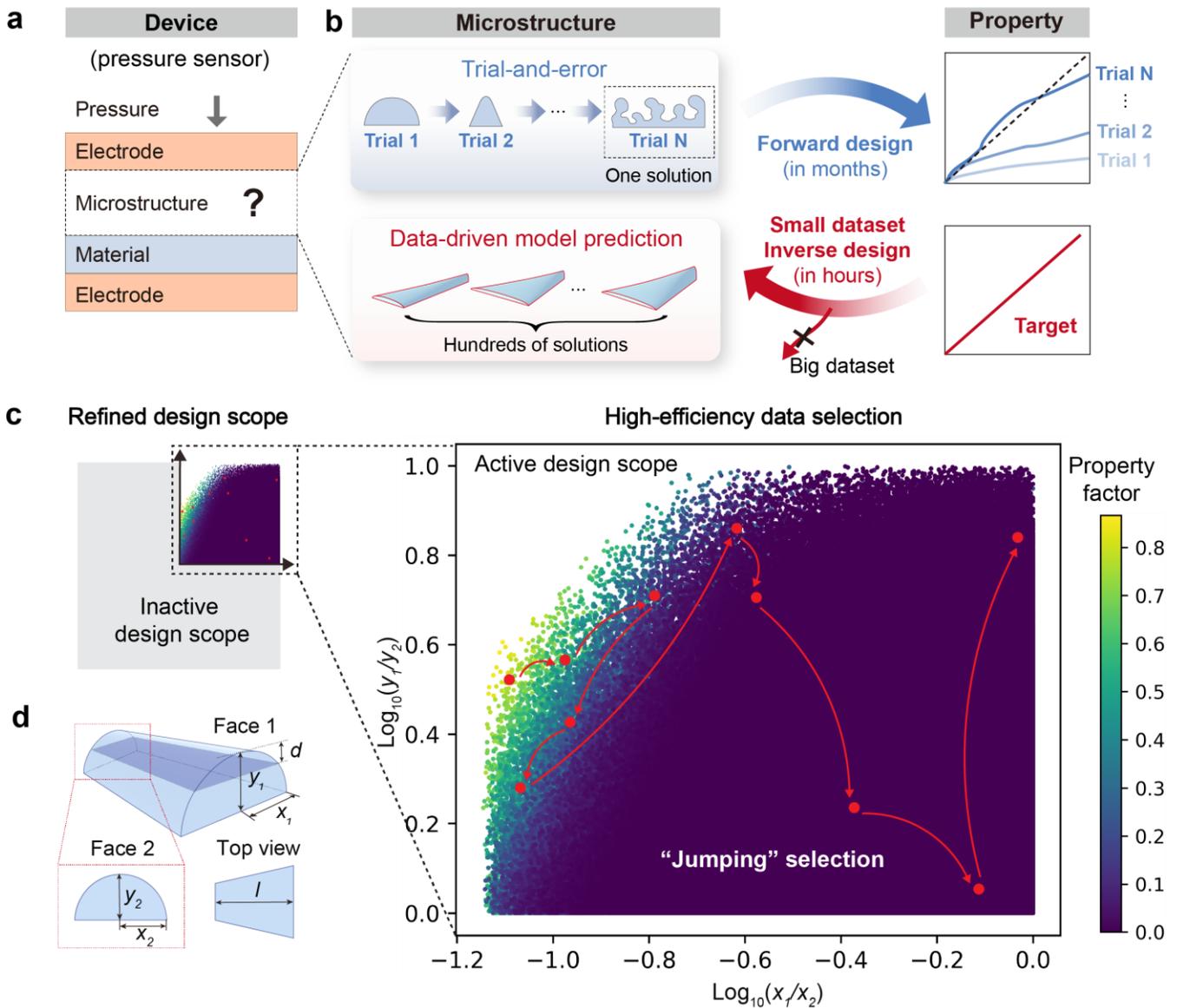

**Figure 1 | Inverse design of artificial skins using a small dataset. a**, A common device structure of flexible pressure sensors. **b**, Comparison between conventional forward design (from structure to property) and the "property to structure" inverse design in this work. **c,** Schematic illustrations of the approaches that release the requirement on the size of dataset. **d**, Microstructure model of this study. All concerned geometric parameters are indicated.

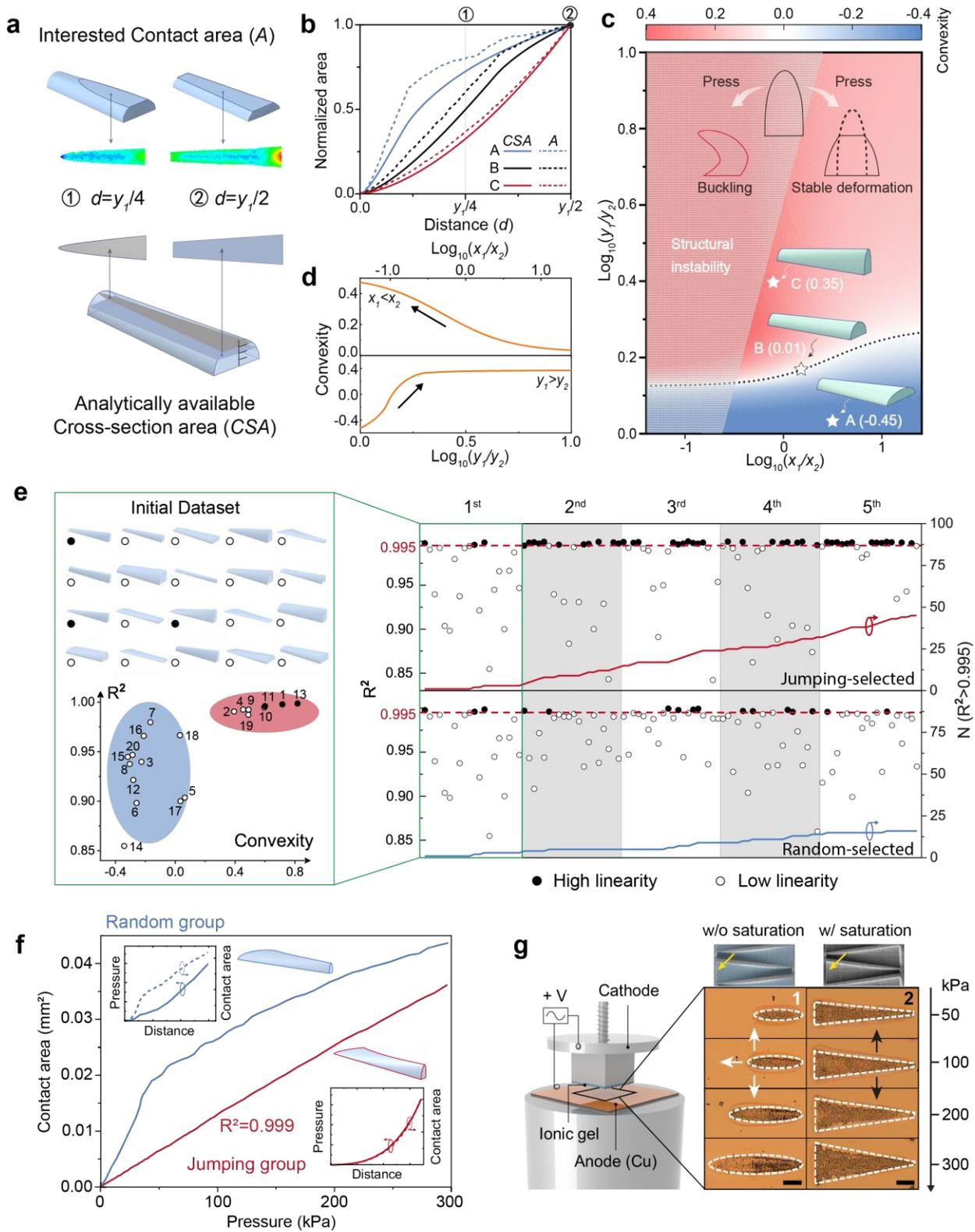

**Figure 2 | Model, design scope, efficiency, and validation of the inverse design. a**, Schematic illustrations of the contact areas (*A*) and cross-section areas (*CSA*) of the same microstructure at compressive-distances (*d*) of $y_1/4$ and $y_1/2$. **b**, *A* and *CSA* as functions of *d* of three representative microstructures, featured as negative (A, convexity=–0.45), near-zero (B, convexity=0.01), and positive convexity (C, convexity=0.35). The corresponding geometric parameters ($\lg(x_1/x_2)$, $\lg(y_1/y_2)$) are (0.5, 0.05), (0.2, 0.17), and (–0.15, 0.4) for case A, B, and C, respectively. **c**, Convexity map calculated by sweeping geometric parameters $x_1/x_2$ and $y_1/y_2$, where the values of $x_2$ and $y_2$ are normalized. Only $y_1>y_2$ is presented here due to structure symmetry. The dashed line draws the boundary between areas of positive and negative convexity. **d**, Average convexity distribution as a function of geometric parameters. **e**, Efficient data collection. Left: twenty random-generated microstructures as the initial dataset, and their linearities with respect to convexities. Right: $R^2$ distributions and accumulative number of eligible geometries ($R^2>0.995$) of the jumping-selected and the random-selected groups in 5 iterations (total 100 cases). **f**, Representative microstructures from the jumping dataset and the random dataset, as well as their calculated *A*-pressure curves. Insets show calculated pressure and contact area with respect to compressive distance. Geometric parameters: Jumping selection: $x_1=10.66$, $y_1=40.34$; $x_2=135.66$, $y_2=10.00$. Random selection: $x_1=10.15$, $y_1=39.42$; $x_2=139.78$, $y_2=37.36$. (**g**) Micrographs of electro-etched patterns for the validation of change in contact area. The structure with a linear response shows both longitudinal and lateral expansion in contact area upon loading, while the control sample show only lateral expansion.

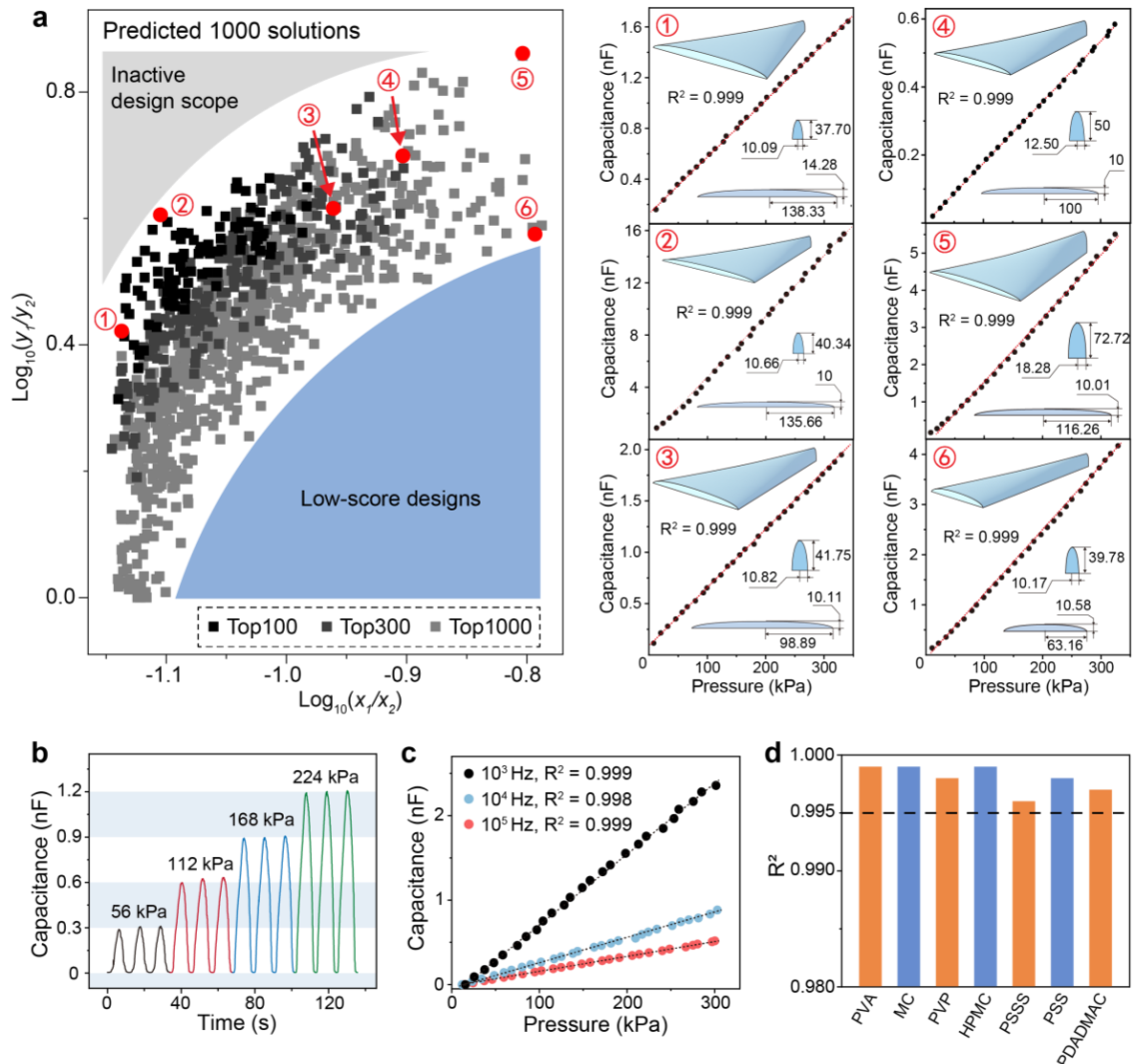

**Figure 3 | Effectiveness and generality for the inverse design of sensors with linear response. a**, Geometric parameter distribution of top 1000 solutions predicted by the ML model. Six microstructures are selected and verified to have linear response over 300 kPa in experiment. The ionic material used is PVA-$H_3PO_4$. **b**, Responses under stepwise loadings of 56, 112, 168, and 224 kPa. **c**, Linear response being maintained under different measurement frequencies. **d**, $R^2$ values of sensors with different ionic gels using a microstructure in panel **a** (no.4), showing high linearity ($R^2$>0.996) for all cases.

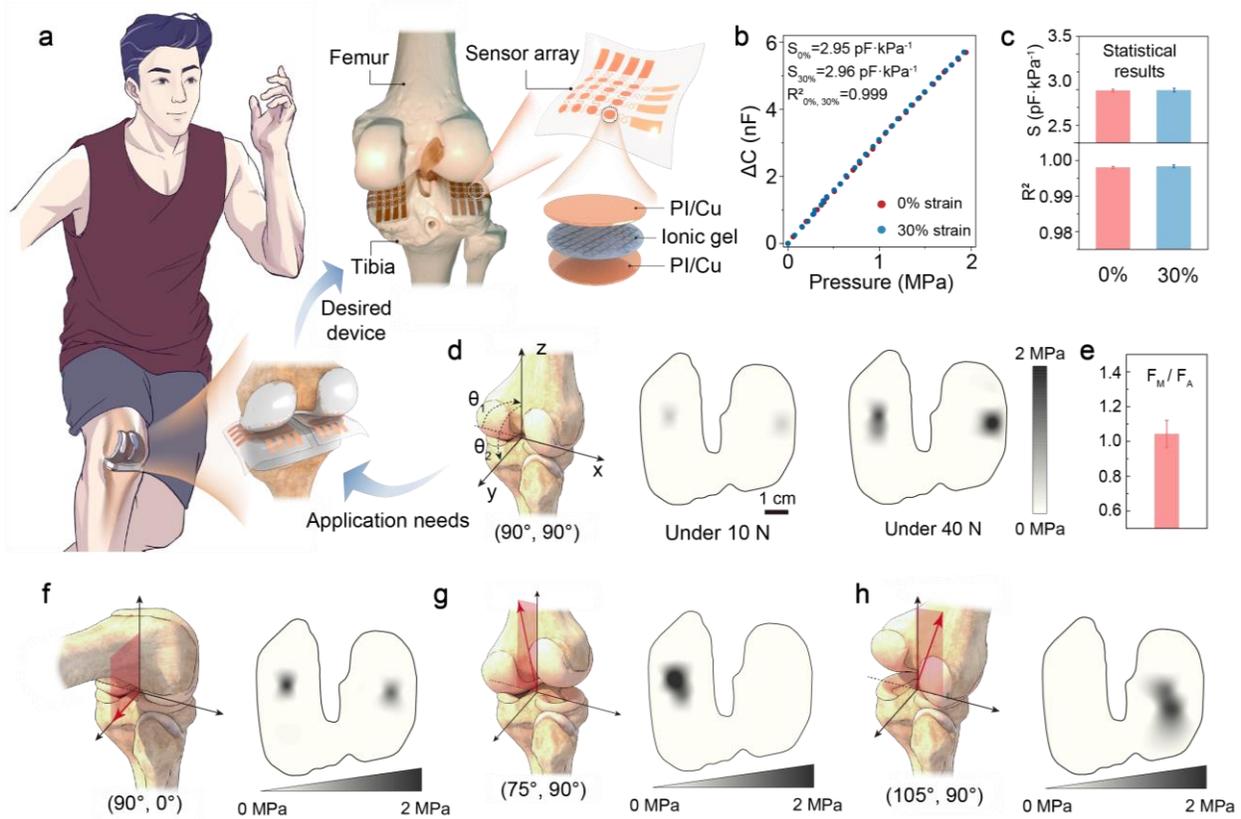

**Figure 4 | Artificial skins designed by inverse design for TKA applications. a**, Schematic illustration for pressure sensing in TKA operation using flexible pressure sensor arrays, which are placed between the pseudo femur and tibia. **b**, Change in capacitance as a function of pressure over 2 MPa for a sensor in non-strained state and under 30% strain. **c**, Statistical results of measured sensitivity and linearity of the sensing units in a sensor array. **d**, Pressure mapping under loads of 10 and 40 N when the pseudo femur is placed vertically. **e**, Measured $F_M/F_A$. The ideal value is 1. **f-h**, Pressure mapping at different contact statuses between the pseudo femur and the tibia.

**Extended data**

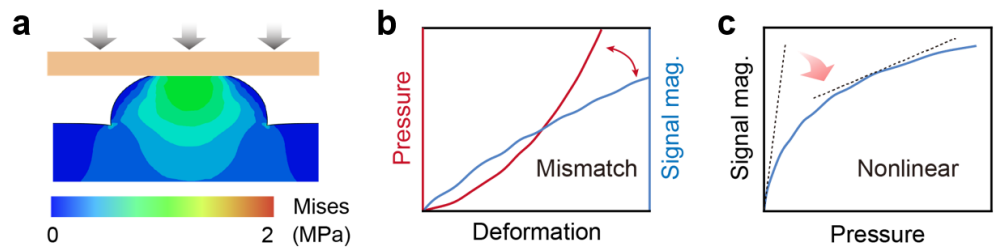

**Extended Data Figure 1. Deformation of an incompressible hemisphere. a,** Simulated deformation of a 2D-hemisphere under pressure. **b,** Simulated curves of input pressure and output signal (contact area) with respect to the deformation of the hemisphere. A clear mismatch is observed. **c**, Nonlinear response curve derived from the stiffening effect, presenting as saturated response.

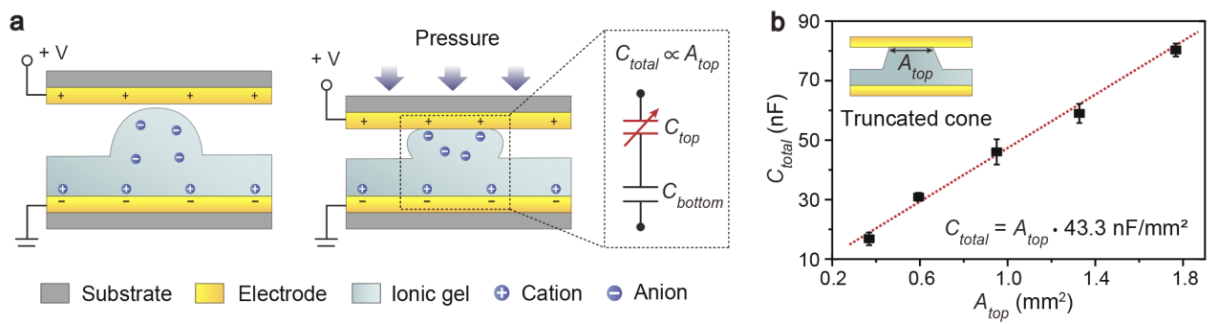

**Extended Data Figure 2. a**, Schematic illustrations of the iontronic pressure sensor and its sensing mechanism. The total capacitance ($C_{total}$) can be approximately expressed as $C_{total} \approx C_{top} \propto A_{top}$ because the contact area at the bottom interface is much larger than the top one ($A_{bottom} \gg A_{top}$). **b**, Experimental results of measured capacitance as a function of contact area of the top interface. PVA-$H_3PO_4$ serves as the ionic layer.

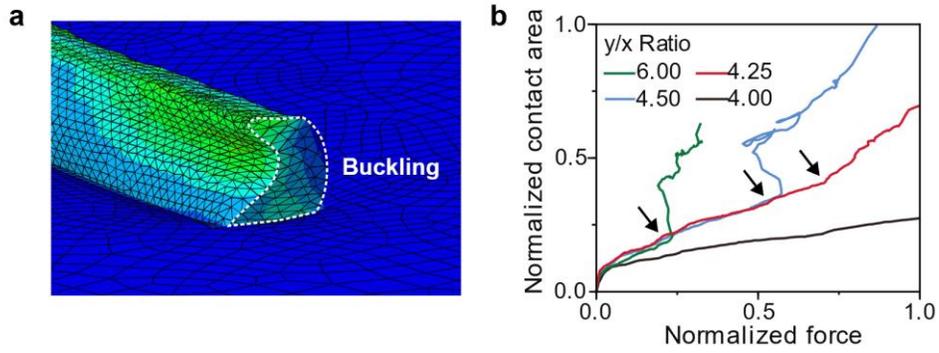

**Extended Data Figure 3. Buckling of a microstructure during compression. a,** Schematic illustration of buckling during compression. **b,** Normalized contact area as a function of normalized force at different *y/x* ratios of 4.00, 4.25, 4.50 and 6.00, assuming $x_1=x_2$, $y_1=y_2$. Buckling occurs for the cases of high *y/x* ratios (4.25, 4.50 and 6.00), which should be avoided.

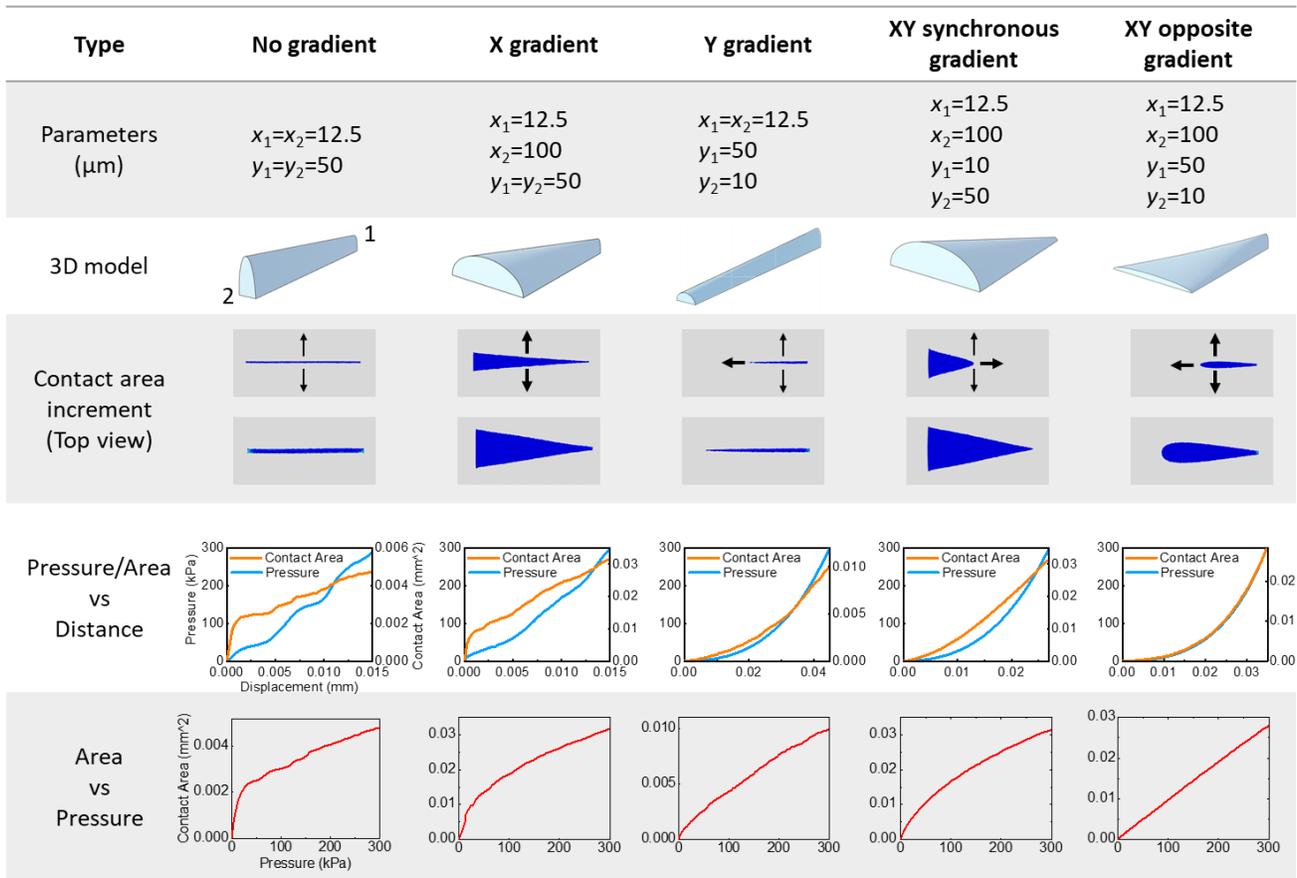

**Extended Data Figure 4.** Correlations among the geometric features, contact area increment, and response curve.

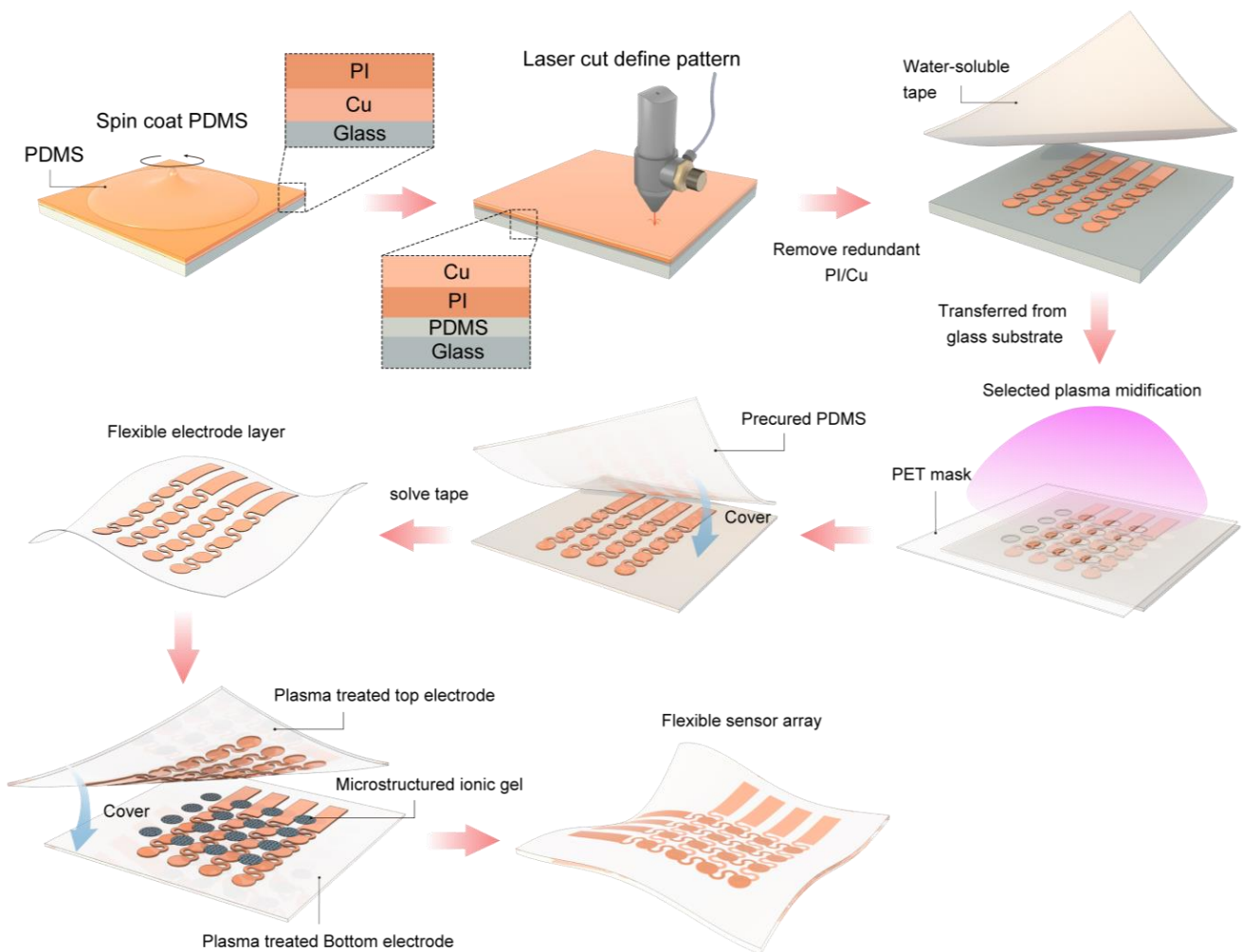

**Extended Data Figure 5.** Flow chart for the fabrication of the sensor array.